\documentclass[aps,prb,twocolumn,nofootinbib]{revtex4-2}
\usepackage{graphicx}
\usepackage{physics}
\usepackage{color}
\usepackage{comment}
\usepackage{siunitx}

\begin{document}

\title{Quantitative Theory for the Amplitude of Fluorescence Quantum Beats from Geminate Triplet-Pair Fusion}

\author{Zachary Rex, B{\'a}ch Xu{\^a}n Ph\d{a}m, Gerald Curran III, Ivan Biaggio}
\affiliation{Department of Physics, Lehigh University, Bethlehem, PA 18018, USA}
 
\date{\today}

\begin{abstract}
We derive the amplitude of  fluorescence quantum beats caused by geminate triplet-exciton fusion from the full set of  parameters that characterize singlet fission and triplet exciton dynamics. We apply the theory  to tetracene and rubrene single crystals.  We find that the amplitude depends on the fission time in tetracene, but does not do so in rubrene, where it is determined  by the dimensionality of triplet transport. Kinetic Monte Carlo simulations  reproduce the experimental data in both materials, for a fission time of the order of 200 ps and isotropic triplet transport in tetracene, and for  a triplet hopping time along the herringbone axis of 200-300 ps and anisotropic transport  in rubrene.  
\end{abstract}

\maketitle
\section{Introduction}

Singlet exciton fission  causes a photoexcited singlet exciton to split into a pair of triplet excitons, with spin conservation requiring them to  be spin-entangled with an overall spin of zero \cite{Merrifield1968}. This can lead to fluorescence quantum beats if the triplet pair has the ability to fuse again into the emissive singlet state.

After pulsed photoexcitation and singlet  fission, the two triplet excitons may separate and later re-encounter, with the possibility of  fusion into a singlet state and photon emission. Because the singlet character of the triplet pair oscillates in time, the fusion probability at the moment of a re-encounter is time-dependent, leading to the observation of quantum beats as  periodic modulations in the time-dependent photoluminescence  \cite{Chabr1981_QuantumBeatsTetracene,Funfschilling1985,Burdett2012,Wolf2018,Biaggio2026}.

In this work, we present a theoretical analysis of the amplitude of  quantum beats as it is  governed by the interplay between triplet transport, triplet fusion, and singlet fission. Two characteristic timescales are particularly important: the singlet fission time, which determines how rapidly the initial singlet excitation converts into an entangled triplet pair, and the anisotropic hopping times of the triplet excitons, which control their  motion through the crystal. 

The relationship between quantum beats and triplet-pair transport has been recognized early on. Theoretical work showed that exciton encounter statistics depend strongly on the dimensionality of the underlying random walk and that repeated encounters become particularly important in systems exhibiting effectively one- or two-dimensional transport \cite{Suna1970}. Subsequently, tetracene quantum beats were  described using a partial random-walk analysis of geminate triplet-pair diffusion \cite{Funfschilling1985}. While these initial studies highlighted the importance of transport in determining the quantum beats resulting from geminate fusion, they were not complete. In particular, they did not take into account  how transport anisotropy and singlet-fission dynamics together govern the quantum-beat amplitude. Here, we provide an analysis that fully incorporates all the relevant physical effects: stochastic and anisotropic triplet transport with different hopping times in each dimension, singlet-fission time,  singlet fission probability, and the effect of multiple subsequent fusion and fission events, together with coherent spin evolution at high magnetic fields.

For quantitative predictions of quantum beats as a function of material parameters, we employ an event-driven kinetic Monte Carlo model that captures the stochastic fission time and motion of triplet excitons, and the oscillatory fusion probability associated with the high-field quantum beat limit. 

This leads to the identification of different regimes where hopping or fission times govern  the magnitude of the observed beat amplitude, and to the    analysis of quantum beats  as a quantitative probe of exciton dynamics, which we confirm by comparison with experiments in tetracene and rubrene.

\section{Theoretical Model}

This work focuses on the initial amplitude of single-frequency quantum-beats obtained in the high magnetic field limit, where transport-induced decoherence \cite{Curran2025TransportDecoherence} is excluded. In this limit, the probability that a spin-coherent entangled triplet pair can undergo fusion upon a re-encounter is modulated by the singlet projection probability 
\begin{equation}
P_{\mathrm{fusion}}(t^*) =
\frac{5}{9}\left(1+\frac{4}{5}\cos(\omega t^*)\right), \label{pFusion}
\end{equation}
where $\omega = 2 \pi f$ is the  beat frequency determined by the energy difference between stationary states in the triplet-pair wavefunction, and $t^*$ is the time elapsed since the last singlet fission event that created the exciton pair. According to this probability function, the maximum quantum beat amplitude obtained by dividing the experimental photoluminescence data by its average non-oscillating trendline \cite{Wolf2018, Curran2024} is equal to 0.8, with an average fusion probability of $P_{\rm fusion} = 5/9$. However, this does not  account for triplet transport and the possibility that a triplet pair undergoes repeated fusion and fission events.

\begin{figure}[t]
\centering
\includegraphics[width=0.98\linewidth]{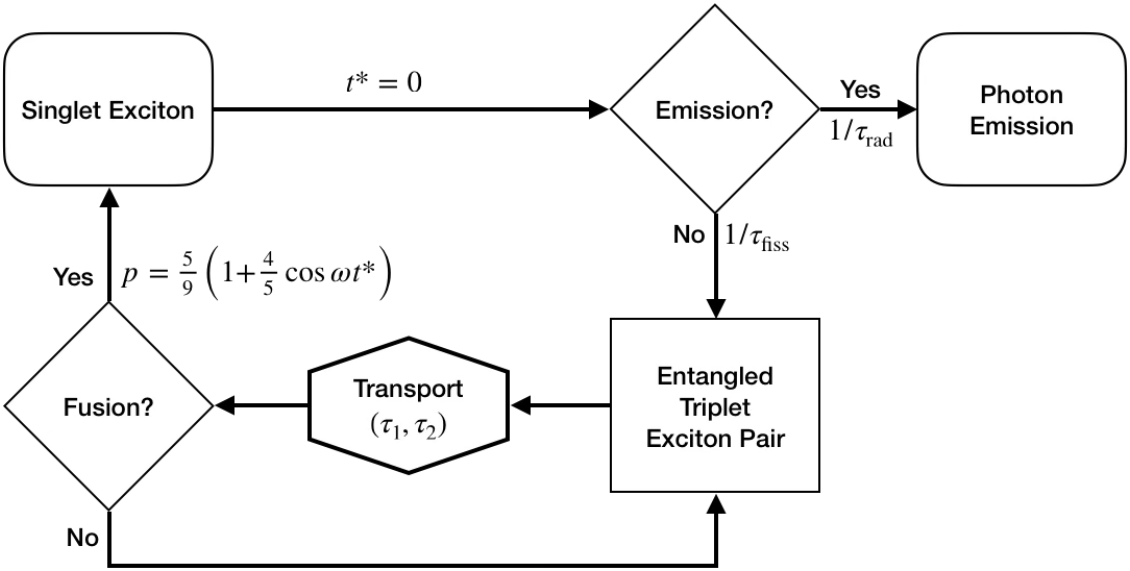}
\caption{ Schematic of the event-driven Monte Carlo simulation. The figure  illustrates the cycle of ``singlet fission -- triplet-pair evolution -- triplet fusion'' that is  interrupted by a final triplet-triplet annihilation (triplet fusion followed by photon emission). Clockwise from the top left: (i) A singlet exciton  decays with emission of a photon or undergoes fission, as determined by drawing from from  exponential distributions defined by the average times $\tau_{\rm rad}$ and $\tau_{\rm fiss}$ of both processes. Fission into a triplet pair is followed by (ii) competing anisotropic triplet hopping (relevant average hopping times $\tau_1$ and $\tau_2$), which (iii) may lead to   a re-encounter, followed by a (iv) fusion attempt governed by $P_{\mathrm{fusion}}(t^*)$, which can result in the re-creation of a singlet exciton and a return to (i) after a time $t^*$. For high fission probability and low photon emission probability the cycle itself runs a large number of times, and the whole simulation is repated for a large number of iterations. The time-dependent probability density of photon emission, $P_\gamma(t)$,  is obtained from a histogram of  all the photon emission times accumulated over all iterations and used to model the experimental time-dependent photoluminescence.}
\label{fig:algorithm}
\end{figure}

To consider the physical events that occur between photoexcitation of a singlet state and the final photon emission (triplet-triplet annihilation), we use a kinetic Monte Carlo model, described in Fig.~\ref{fig:algorithm}. By running the model over multiple iterations, we obtain the time-dependent photon emission probability density, which corresponds to the time-dependent photoluminescence emitted after short-pulse excitation.

\begin{figure}[t]
\centering
\includegraphics[width=0.98\linewidth]{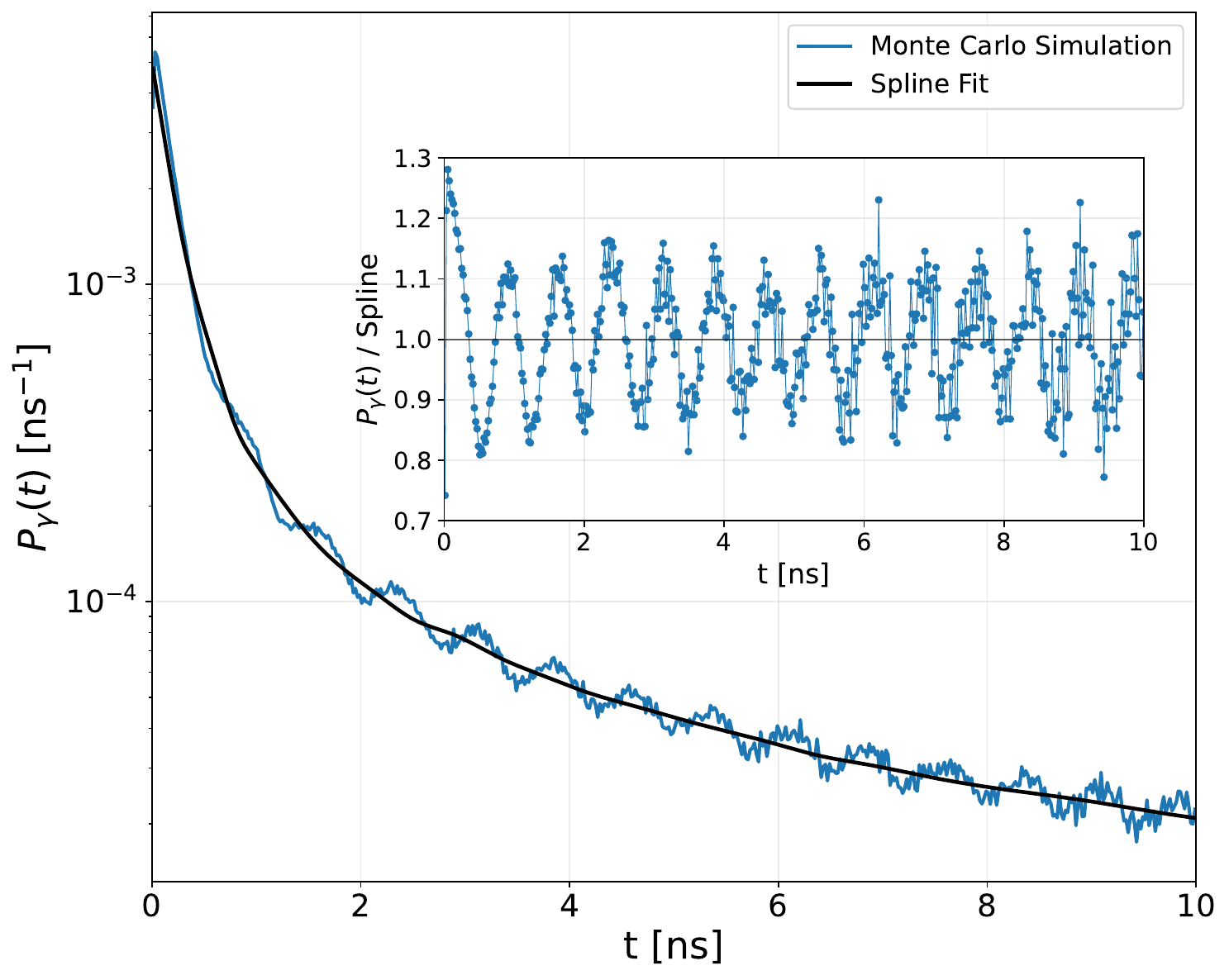}
\caption{Time-dependent photon emission probability density $P_\gamma(t)$ obtained from a Monte-Carlo simulation with $\tau_{1} = 3$~ps, $\tau_{2}= 250$~ps, $f = 1.34$~GHz, $\tau_{\rm rad} = 15.2$~ns, $\tau_{\rm fiss}= 10$~ps ($p_{\rm fiss} = 0.9993$), $N_{\rm iter} = $~\num{2.0e7}. The solid black curve is the average trendline obtained via a spline fit. The inset shows the quantum beats extracted by dividing $P_\gamma(t)$ by this average trendline.}
\label{fig:tdpl}
\end{figure}

Table~\ref{tab:mc_parameters} lists the parameters used in our model, and an example of the result of the Monte Carlo procedure is shown in Fig.~\ref{fig:tdpl}.

\begin{table}
\centering
\caption{Material parameters used in the theoretical model. The Monte Carlo simulation allows individual parameters to be varied independently to investigate their influence on the quantum beat amplitude. The ``Rubrene'' and ``Tetracene'' columns give the parameter values obtained by comparing the model with experimental data. Values in parentheses are derived from the other parameters. The physical meanings of each parameter are explained in the text.
}

\begin{tabular*}{\linewidth}{@{\extracolsep{\fill}}llll}
\hline\hline
Parameter &  Rubrene & Tetracene \\
\hline
$\tau_1$                                  & 3 ps          & 3 ps \\
$\tau_2$                                    & 250 ps    & 5 ps \\
$p_{\mathrm{f.a.}}$            & 0.3             & 0.3 \\
$\tau_{\mathrm{fiss}}$                          & 10 ps     & 200 ps \\
$\tau_{\mathrm{rad}}$                       & 15.2 ns   & 5 ns \\
$p_{\mathrm{fiss}}$                             & (0.9993)  & (0.962) \\
\hline\hline
\end{tabular*}
\label{tab:mc_parameters}
\end{table}

\begin{figure}[t]
\centering
\includegraphics[width=1\linewidth]{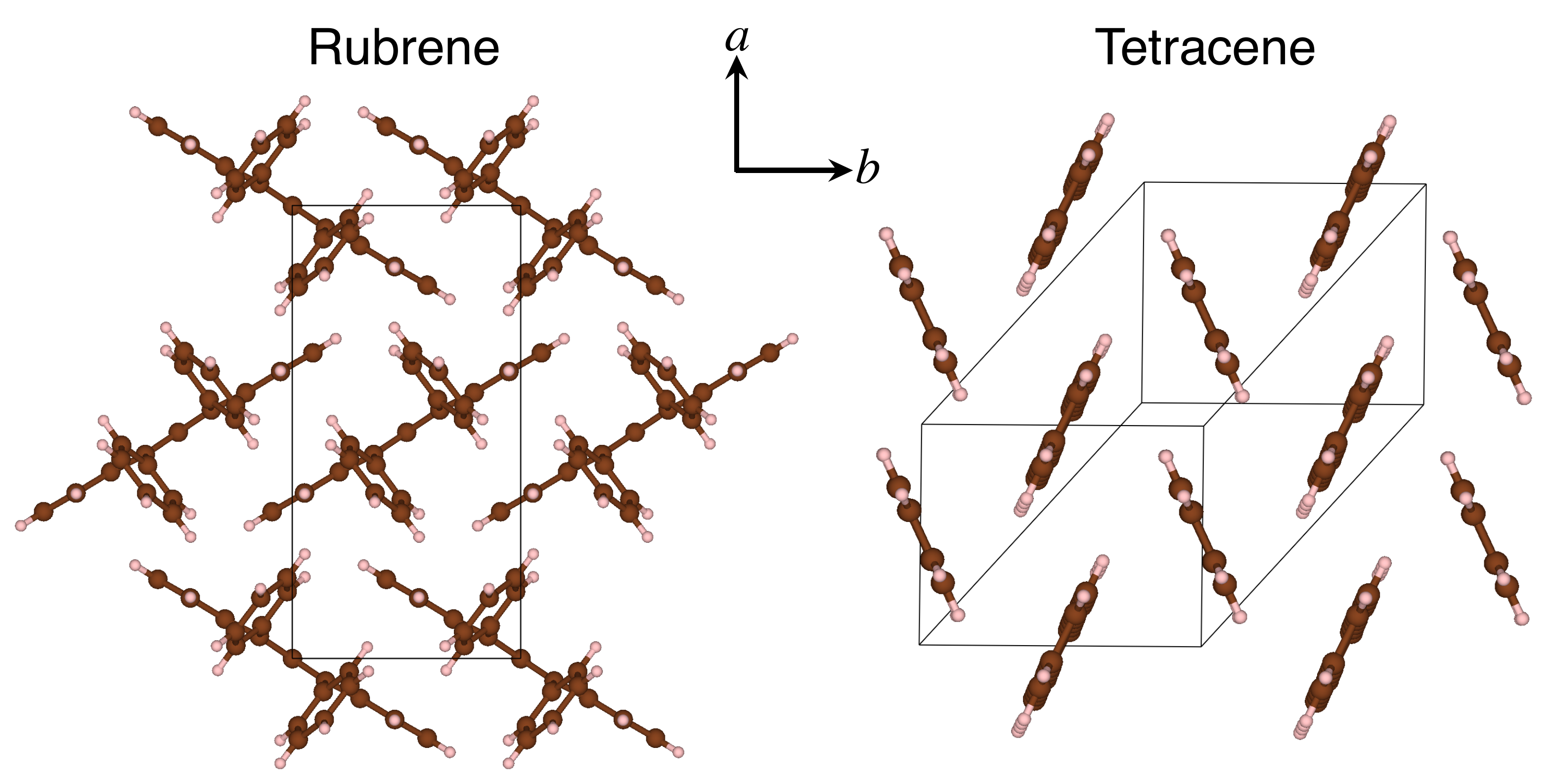}
\caption{Crystal structures of rubrene (left) and  tetracene (right) in the molecular plane where triplet-exciton diffusion primarily  occurs, illustrating the different molecular packing arrangements that govern triplet exciton transport. The unit cells and crystallographic axes are indicated for both materials, in the convention of Refs.~\onlinecite{Wolf2018,Curran2024} for rubrene and in the convention of Refs.~\onlinecite{Robertson1961Tetracene,Akselrod2014}   for tetracene. The shortest triplet exciton hopping time $\tau_1$ used in our model applies to transport along the $b$ direction. In rubrene, the molecular arrangement favors strongly anisotropic transport, with rapid hopping along the molecular stacks in the $b$-direction and substantially slower hopping along the $a$-axis \cite{Curran2024}. In contrast, the packing of tetracene allows for  nearly isotropic triplet transport within the  molecular plane shown \cite{Akselrod2014}.}
\label{rubrene_tetracene_packing}
\end{figure}

The Monte Carlo implementation (Fig.~\ref{fig:algorithm}) is based on  constant probabilities per unit time for all events, and it simulates spatial random walks for both triplet excitons in the pair. Hopping time and direction are obtained by drawing random times  from  exponential distributions with mean values $\tau_{i}$ for each spatial dimension, with the smallest time determining the next hopping event and its direction (only the two shortest hopping times, $\tau_1$ and $\tau_2$ are relevant for the results presented here). The molecular arrangement in the plane where the fastest triplet exciton diffusion occurs is shown for both rubrene and tetracene in Fig.~\ref{rubrene_tetracene_packing}.

Similar decision-making is implemented at every step, including fission instead of radiative decay after fusion, or fusion instead of hopping on the occasion of a re-encounter.  Fission and radiative decay are treated as the only two  possibilities after a successful fusion event, and parameterized through average time constants $\tau_{\rm rad}$ and $\tau_{\rm fiss}$, corresponding to the inverse of  the probability per unit time that either event occurs. Thus, in this model radiative decay effectively corresponds to triplet-triplet annihilation, and the same parameters can also be used to model annihilation probability in cases where triplet-triplet interaction cannot result in an emissive singlet state,

When evaluating the effect of different fission probabilities, we use a fixed fission time and a derived ad-hoc  time $\tau_{\rm rad} = \tau_{\rm fiss}/(p_{\rm fiss}^{-1} - 1)$.
We also include a phenomenological  probability that a fusion attempt is initiated, $p_{\rm f.a.} = 0.3$, to account for the fact that initiating fusion would compete with the triplet-pair separating again, but this choice does not have a relevant effect on the results. Throughout, the time after which every event occurs is recorded and accumulated. Whenever the chain of decision-making leads to a photon emission event before a maximum time is reached, the  photon emission time is stored. 

Repeating the procedure of Fig.~\ref{fig:algorithm} for a large number of iterations produces a distribution of photon emission times that are assigned to bins of duration $\Delta t$. The  number of photons $n_\gamma(t)$ in each bin after $N_{\rm iter}$ iterations then delivers the photon emission probability density per unit time, $P_\gamma(t) = n_\gamma(t)/\Delta t/N_{\rm iter}$, which we use to model the experimental time-dependent photoluminescence. Quantum beats are extracted  by taking the ratio between $P_\gamma(t)$ and its average trendline,  the same  method used for  experimental data \cite{Wolf2018,Curran2024}.

\begin{figure*}[t]
\centering
\includegraphics[width=\textwidth]{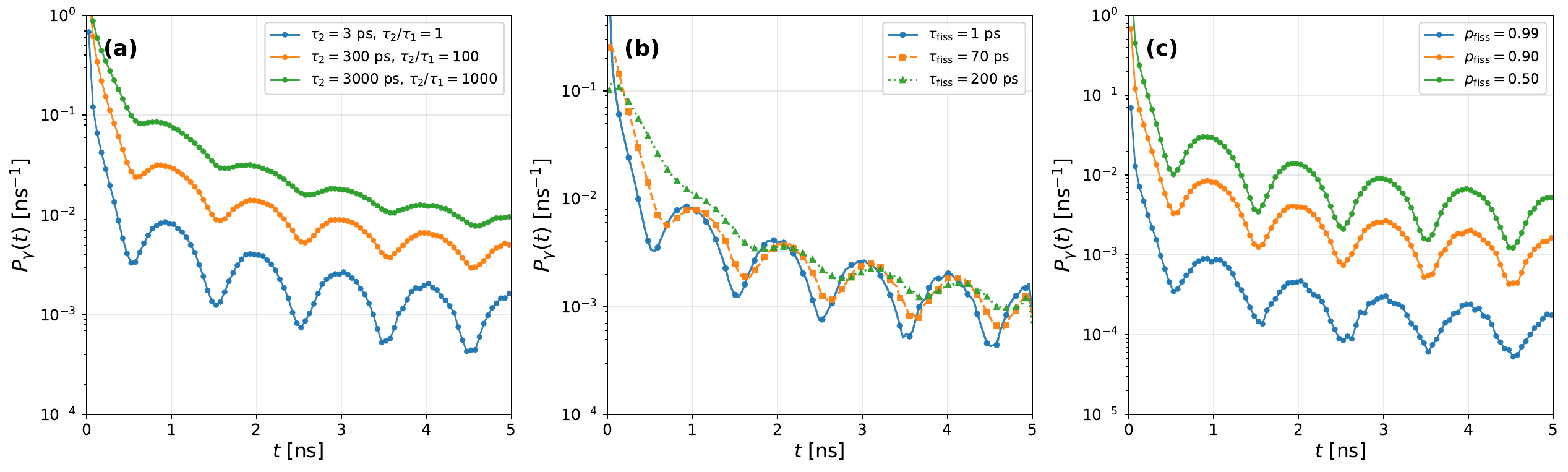}
\caption{Time-dependent photon emission probability density $P_{\gamma}(t)$  for different values of key parameters. ($a$): for different  hopping time $\tau_{2}$ with  $\tau_{\rm fiss}$ = 1 ps and $p_{\rm fiss}$ = 0.90, ($b$): for different fission times $\tau_{\mathrm{fiss}}$, with $\tau_{2}$ = 3 ps and $p_{\rm fiss}$ = 0.90, and ($c$): for different singlet fission probability $p_{\mathrm{fiss}}$, with $\tau_{2}$ = 3 ps and $\tau_{\rm fiss}$ = 1 ps. In each panel, only the indicated parameter is varied. In all cases, $\tau_{1}$ = 3 ps and $N_{\rm iter}$ = \num{2.0e7}.}
\label{fig:tdpl_panel}
\end{figure*}

The Monte Carlo simulation also delivers a singlet fission efficiency $\eta_{SF} = 1 - N_{\gamma}/N_{\rm iter}$, where  $N_{\gamma}$ is the number of iterations that resulted in triplet-triplet annihilation. This quantity  gives the  survival probability of the triplet pair after the maximum time set in the simulation, and converges towards a constant efficiency $\eta_{SF}$ as the maximum time grows. For all the results presented here, we use a maximum time for the Monte Carlo simulation of 10 ns. $\eta_{SF}$ does not change significantly when choosing larger simulation durations because the probability of a re-encounter drops quickly, approximately with the inverse of time, $\propto t^{-1}$, for a 2D random walk \cite{Wolf2021}. We note that the singlet fission efficiency obtained in this way  applies in general to any material where  triplet-triplet annihilation can occur upon a re-encounter, not necessarily through fusion and radiative decay.

\section{Predicted quantum beats}

The key dynamic parameters that influence the  photoluminescence quantum beats are the fission efficiency---as represented by either  fission probability $p_{\mathrm{fiss}}$ and fission time, or by decay time $\tau_{\rm rad}$ and fission time---and the two fastest hopping times $\tau_1$ and $\tau_2$. For both rubrene and tetracene $\tau_3$ is much longer (on the microsecond time scale for rubrene \cite{Wolf2021}) and it does not affect the results.

In the following, we determine how each of these parameters influences the calculated  photon emission probability density $P_{\gamma}(t)$ (corresponding to the experimental time-dependent photoluminescence), the  quantum beat amplitude, and the singlet-fission efficiency. A first set of results is shown in Fig.~\ref{fig:tdpl_panel}.

In general, $P_\gamma(t)$ exhibits a common structure consisting of an initial rapid decay followed by a slower tail that reflects the decreasing likelihood of re-encounter during the random walk \cite{Wolf2021}. For our choice of negligible hopping in the third dimension on the investigated time scale, this slower tail  tends to a power-law decay with the expected $\sim -1$ exponent  \cite{Wolf2021}. This  trend is then modulated by the quantum beat oscillations that originate from the time dependence of the singlet projection probability.
Several observations can be made about these results. 

The average emission probability  clearly increases with the slower hopping time $\tau_2$ or the slower fission probability (Fig.~\ref{fig:tdpl_panel}$a$ and Fig.~\ref{fig:tdpl_panel}$c$), while it is less affected by the fission time (Fig.~\ref{fig:tdpl_panel}$b$). This can be ascribed to the fact that an increase in $\tau_2$ leads to a longer confinement of the diffusion in the triplet exciton pair to one dimension, which increases the probability of a triplet-pair re-encounter and the number of   repeated fusion and fission events, which  ultimately increases the probability of annihilation with photon emission.  On the other hand, the  overall photon emission probability remains determined by the re-encounter probability and the constant fission probability assumed in Fig.~\ref{fig:tdpl_panel}$b$.

However, and in contrast to the average emission probability, the quantum beat amplitude is not significantly affected by the fission probability. Instead, it critically depends on both the slower hopping time $\tau_2$ and the fission time: an increase in the value of either of these parameters causes an  attenuation of quantum beat amplitude because of dephasing of the spin-wavefunction of different triplet pairs.

\begin{figure*}[t]
\centering
\includegraphics[width=\textwidth]{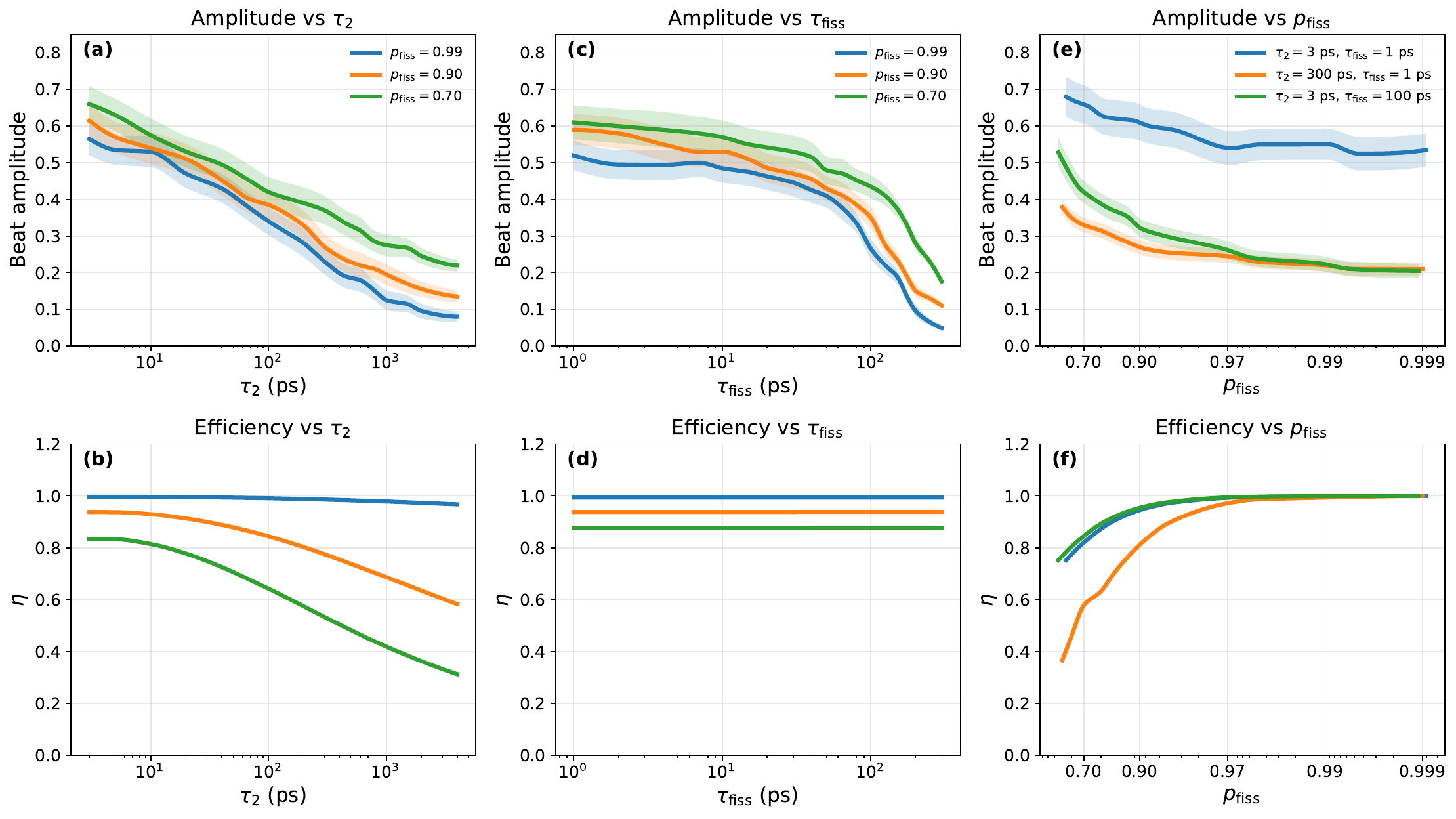}
\caption{Quantum beat amplitude (top row) and singlet fission efficiency (bottom row). (a,b): as a function of $\tau_{2}$, with $\tau_{\rm fiss}$ = 1 ps and the three different $p_{\rm fiss}$ specified in the figure. (c,d):  as a function of $\tau_{\mathrm{fiss}}$, with $\tau_{2}$ = 3 ps, and the different $p_{\rm fiss}$ specified in the figure. (e,f):  as a function of $p_{\mathrm{fiss}}$, with the three different choices of  $\tau_{2}$ and $\tau_{\rm fiss}$ specified in the figure. For all cases, $\tau_{1}$ = 3 ps and $N_{\rm iter}$ = \num{2.0e7}.}
\label{fig:summary_panel}
\end{figure*}

To better quantify how these material properties affect singlet fission and the coherence in the triplet-pair population, we extract the quantum beat amplitude and the singlet fission efficiency from the same Monte Carlo simulation that lead to Fig.~\ref{fig:tdpl_panel}.  These results are shown in Fig.~\ref{fig:summary_panel}.

The quantum beat amplitude continually decreases when increasing  either  the $\tau_2$ hopping time or the fission time at constant fission probability, even though the decay as a function of the $\tau_2$ hopping time is more gradual (Fig.~\ref{fig:summary_panel}$a$ and Fig.~\ref{fig:summary_panel}$c$).  The root reason for the amplitude decrease is that uncertainty in the exact time when either process occurs can establish  random phase-shifts in the time-dependent wavefunction of stationary states,  which causes  a population-level attenuation of the oscillating fusion probability due to quantum interference. But this happens through two distinct mechanisms.

The slower hopping time $\tau_2$ affects the statistics of re-encounters through the way in which it can limit triplet diffusion to one dimension, where a larger re-encounter probability can lead to multiple fusion and fission events in the same triplet-pair. An increase in $\tau_2$ at a constant (short) fission time,  creates a larger uncertainty in the duration of each one-dimensional transport period, with a corresponding uncertainty in the timing of the last fission event before the two triplet excitons can separate by two-dimensional diffusion. This period of accumulated one-dimensional triplet-exciton  diffusion causes random time offsets in the time-dependent singlet-projection probability function of each triplet pair, even for a situation where the singlet-fission time was instantaneous.

On the other hand, a longer fission time also leads to a larger uncertainty in the timing of each fission event. This again leads to a randomized time-offset  in the time-dependent singlet-projection probability function of each triplet pair, which happens also in the  case of isotropic two-dimensional diffusion, where both triplet-exciton hopping times are similar.

In summary,   while $\tau_{2}$ controls the number of times a re-encounter event occurs, thus re-setting (through a fusion and re-fission event) the time when the triplet-pair spin wavefunction starts to evolve away from an overall singlet state, $\tau_{\mathrm{fiss}}$ directly governs the degree by which each individual fusion and re-fission event degrades phase coherence by itself.

\begin{figure*}[t]
\centering
\includegraphics[width=\textwidth]{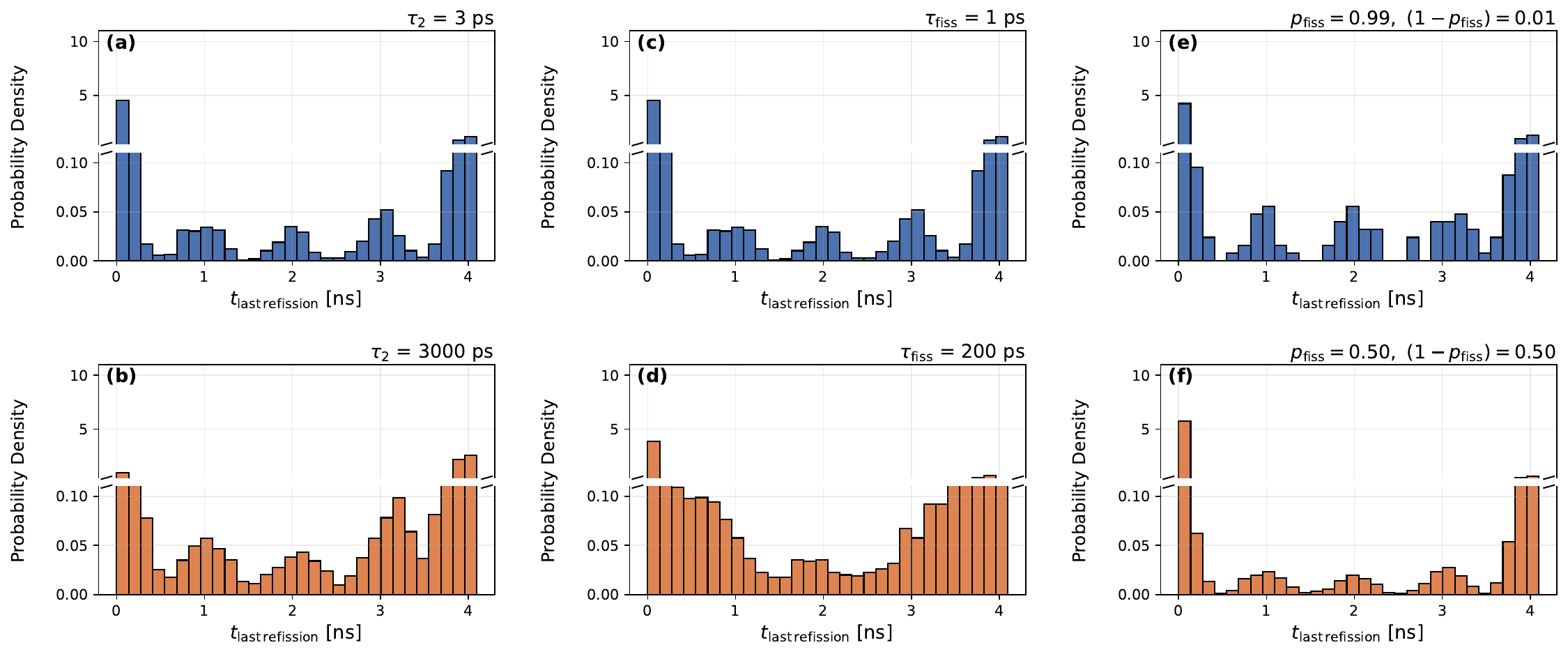}
\caption{Distributions of the last fission time
for photon emission events in the time interval between $t$ = 3.8 ns and $t$ = 4.2 ns. (a,b): for the indicated values of  $\tau_{2}$, with $\tau_{\rm fiss}$ = 1 ps and $p_{\rm fiss}$ = 0.9, (c,d). for the indicated values of $\tau_{\mathrm{fiss}}$, with $\tau_{2}$ = 3 ps and $p_{\rm fiss}$ = 0.9.  (e,f): for the indicated values of $p_{\mathrm{fiss}}$, with $\tau_{2}$ = 3 ps and $\tau_{\rm fiss}$ = 1 ps. For all simulations, $\tau_{1}$ = 3 ps and $N_{\rm iter}$ = \num{2.0e7}.}
\label{fig:last_reset_panel}
\end{figure*}

When analyzing the problem in terms of a fission probability $p_{\mathrm{fiss}}$ (Fig.~\ref{fig:summary_panel}$e$), one sees that  its increase only leads to a less pronounced reduction in beat amplitude by the fact that it controls how many fusion and re-fission cycles can happen. But this reduction is limited, even for longer times, because the time-dependent probability of each successive cycle of fusion and re-fission to occur is informed by previous cycles in such a way that information about the timing of the first fission is carried forward in time (see the discussion around Fig.~\ref{fig:last_reset_panel}, below).  

The same parameters influence singlet fission efficiency in a different way. First, one sees that it naturally drops rapidly at lower fission probabilities, which directly  increase the triplet-triplet annihilation probability (Fig.~\ref{fig:summary_panel}$f$)). But the singlet fission efficiency also  decreases (especially for lower fission probabilities that increase the chance for annihilation)  whenever the triplet pair is confined to one-dimensional transport for a longer time (Fig.~\ref{fig:summary_panel}$b$)), with the multiple re-encounters then again increasing the  probability of triplet-triplet annihilation.  On the other hand, as expected, singlet fission time does not affect fission efficiency at all.

Further insight into the behaviors described above can be obtained by examining the statistical distribution of the   last re-fission event prior to photon emission, shown in Fig.~\ref{fig:last_reset_panel}. The histograms in this figure plot the time distribution of the last fusion and re-fission event that occurred before a photon emission event at 4~ns after the initial fission event.

These distributions demonstrate that the beat amplitude is generally set early in the process (on the time scale of $\tau_2$ and $\tau_{\rm fiss}$) and that the system retains a  memory of the first fission event after photoexcitation, which ultimately maintains a degree of coherence in the triplet-pair population. This is seen by the fact that the distributions in Fig.~\ref{fig:last_reset_panel}   exhibit an oscillatory structure at the quantum beat frequency, mirroring  the fluctuations in the underlying singlet-projection probability that was set during the first fission event (at time 0 in this figure). Essentially,  the time when a previous  fission event occurs affects the time when the next fusion and re-fission  even can occur, transmitting the information forward in time, which then causes neglibible decrease in beat amplitude once the triplet excitons starts diffusion in two dimensions.

\section{Application to rubrene and tetracene}

The results presented thus far establish a general framework for understanding how stochastic transport, singlet fission dynamics, and recombination probability govern the observable quantum beat signal.
They  show that the observable quantum beat amplitude is governed by the cumulative effect of stochastic reset events during the evolution of the  triplet-pair spin wavefunction. Transport controls how often re-encounters occur, singlet fission time determines how long coherent evolution persists between successive  fusion and re-fission events, and the fission probability sets the likelihood that the system terminates in triplet-triplet annihilation (and photon emission) after fusion. 

Most importantly, the results  show how the quantum beat amplitude can be predicted.  We  confirmed the validity of this framework by comparing with the experimental results in  rubrene and tetracene crystals, also proving that quantum beat measurements can be used to determine the values of important parameters, such as the slower hopping time $\tau_2$ in rubrene, and the fission time $\tau_{\rm fiss}$ in tetracene.

In our experiments, we used 150 fs pulses at a wavelength of 515 nm to photoexcite an initial singlet exciton population at a density of the order of $10^{21}$ m$^{-3}$ near the surface of the crystals, and measured the resulting time-dependent photoluminescence decay. 
This singlet exciton population density is low enough to avoid the possibility of non-geminate fusion of triplet excitons from different fission events, and to ensure that virtualy all of the detected photons originate from geminate triplet-triplet interaction \cite{Wolf2021}. 
In addition, we note that diffusion of triplet excitons towards the surface is negligible on a nanosecond time scale (triplet exciton hopping time in that direction for rubrene is of the order of a microsecond \cite{Wolf2021}), which combined with a sufficiently large absorption length of 515 nm light (several micrometers in rubrene \cite{Irkhin2012Rubrene}), guarantees that the vast majority of excitons never interacts with the surface, and that exciton diffusion is only relevant in directions parallel to  the crystal surface and is effectively 2-dimensional in the $ab$ plane (See Fig.~\ref{rubrene_tetracene_packing}).

All experiments  were performed in a high field limit with a magnetic field  1.5 T for both rubrene and tetracene single crystals. The magnetic field direction was tuned to obtain different quantum beat frequencies, while any transport-induced dephasing \cite{Curran2024,Curran2025TransportDecoherence} was avoided by keeping the field parallel to crystal symmetry planes  \cite{Curran2024,Curran2025TransportDecoherence}. This  ensured that Equation \ref{pFusion} accurately describes the time-dependent fusion probability of the triplet pair within a time interval where spin-relaxation does not occur, and that it is appropriate to model the initial quantum beat signal.

\begin{figure}[t]
\centering
\includegraphics[width=\linewidth]{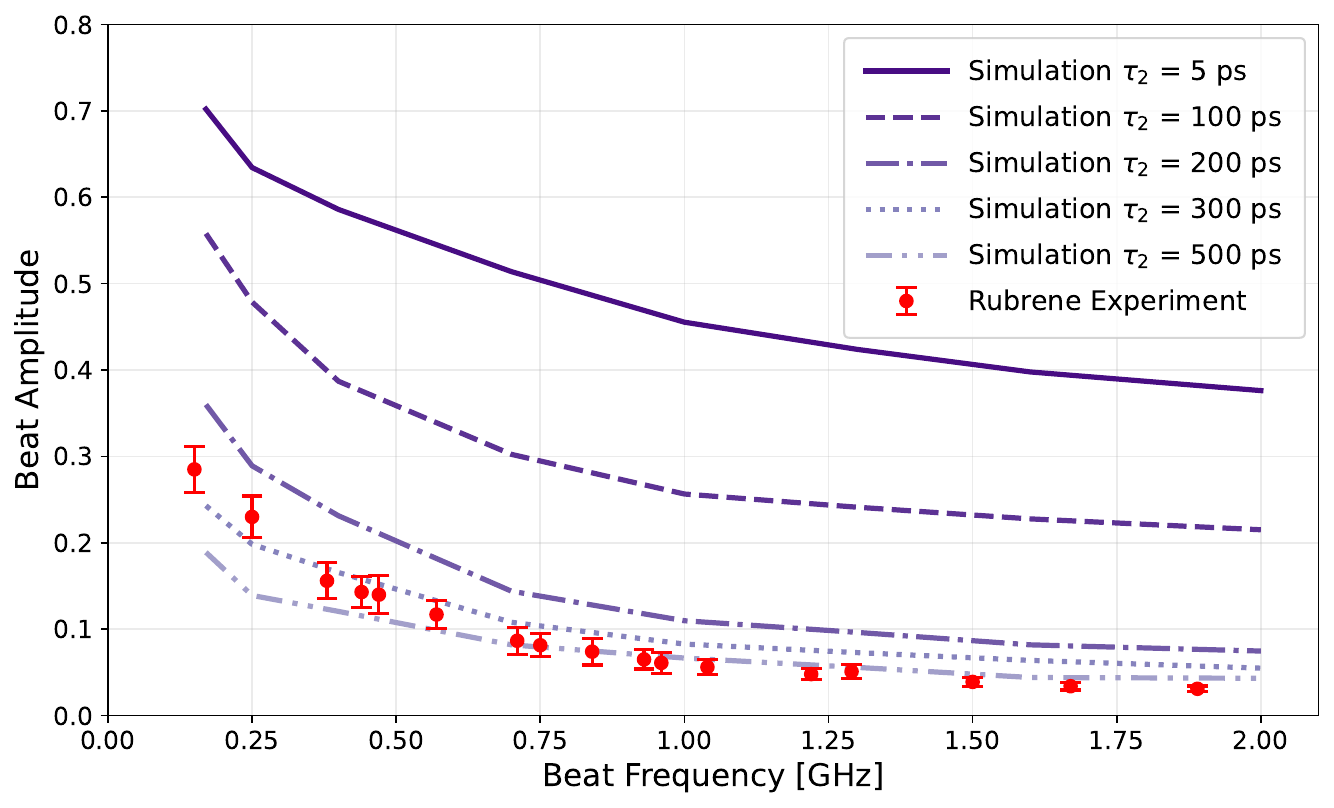}
\caption{Quantum beat amplitude as a function of beat frequency with experimental data for rubrene (red data points), compared to theoretical expectations (purple curves) for different  two-dimensional hopping times $\tau_{2}$. The other parameters are  listed in Table~\ref{tab:mc_parameters} and are $\tau_{1}$ = 3 ps \cite{Irkhin2011}, $\tau_{\rm fiss}$ = 10 ps\cite{Finton2019}, $\tau_{\rm rad}$ = 15.2 ns\cite{Finton2019} ($p_{\rm fiss}$ = 0.9993), $N_{\rm  iter}$ = \num{1.0e8}.
}
\label{fig:rubrene_amp}
\end{figure}

Fig.~\ref{fig:rubrene_amp} shows the experimental data for the quantum beat amplitude in rubrene, together with the prediction form our theoretical model for different values of the  hopping time between inequivalent sites, $\tau_{2}$, and the parameters in Table~\ref{tab:mc_parameters}.

Both experimental data and theoretical predictions exhibit a monotonic decrease in amplitude with increasing frequency, consistent with the expectation that faster oscillations are more sensitive to phase disruption. The simulation curves reproduce this trend and show that the amplitude is strongly controlled by the accumulation of stochastic re-fission events that is affected by  the $\tau_{2}$ hopping time,.
By matching the experimental data to the simulation, we identify a range of $\tau_{2}$ values, 200-300 ps, that capture the observed behavior. This estimate is consistent with a previous estimate obtained with a different method \cite{Curran2024,Curran2025Erratum}.

\begin{figure}[t]
\centering
\includegraphics[width=\linewidth]{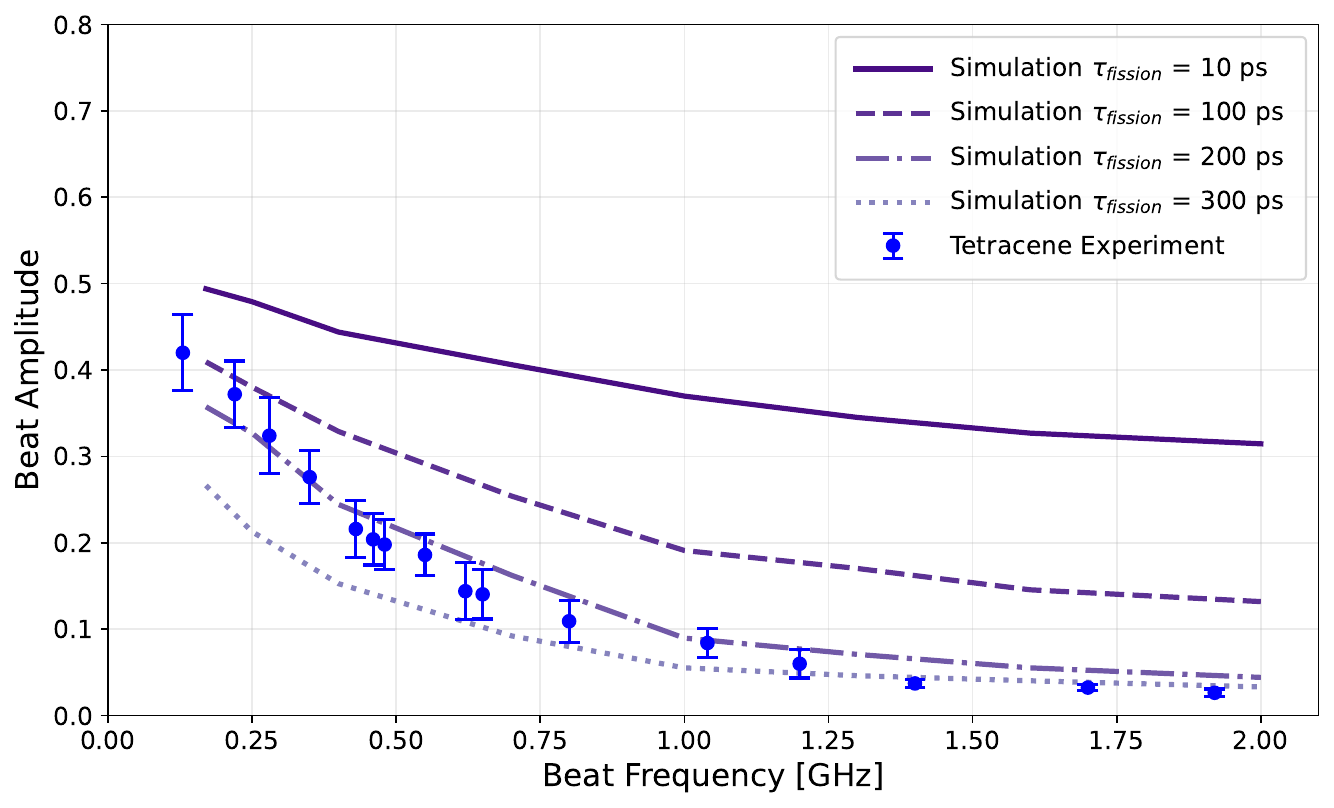}
\caption{ Quantum beat amplitude as a function of beat frequency. Experimental data for tetracene (blue data points) compared to the theoretical prediction (purple curves) for different values of the fission time $\tau_{\rm fiss}$. The other parameters are $\tau_{1}$ = 3 ps, $\tau_{2}$ = 5 ps \cite{Wan2015_CooperativeExcitonTransport,Akselrod2014}, $\tau_{\rm rad}$ = 5.0 ns \cite{Thompson2015_MagneticFieldSF,Lim2004Exciton}, and  $N_{\rm iter} = \num{1.0e8}$. Here, $p_{\rm fiss}$ depends on $\tau_{\rm fiss}$ as $p_{\rm fiss} = (\tau_{\rm fiss}/\tau_{\rm rad} + 1)^{-1}$.}
\label{fig:tetracene_amp}
\end{figure}

Fig.~\ref{fig:tetracene_amp} shows the same results for tetracene, for which  we plot the quantum beat amplitude as a function of the  singlet fission time $\tau_{\mathrm{fiss}}$ while keeping the transport parameters fixed at the values given in Table~\ref{tab:mc_parameters}.

The experimental trend is again a monotonic decrease of beat amplitude with frequency, for both experiment and theoretical prediction. The best agreement obtained between theory and experiment is found for fission times between 150 and 250 ps, consistent with independently measured values for tetracene \cite{Smith1972Tetracene,Alfano1973,Fleming1977Tetracene,Klein1978TripletPairs,Burdett2013,Wilson2013TetraceneFission,Piland2015Tetracene,Felter2019_SingletFissionCrystals}. In both Fig.~\ref{fig:rubrene_amp} and Fig.~\ref{fig:tetracene_amp}, the experimental beat amplitudes at high beat frequencies fall beneath the simulation curve it best matches at lower frequencies.  This  is likely an artifact caused by lower signal-to-noise ratio at high frequencies, where the beat amplitude becomes comparable to the noise of the photoluminescence signal.

\begin{figure}[t]
\centering
\includegraphics[width=\linewidth]{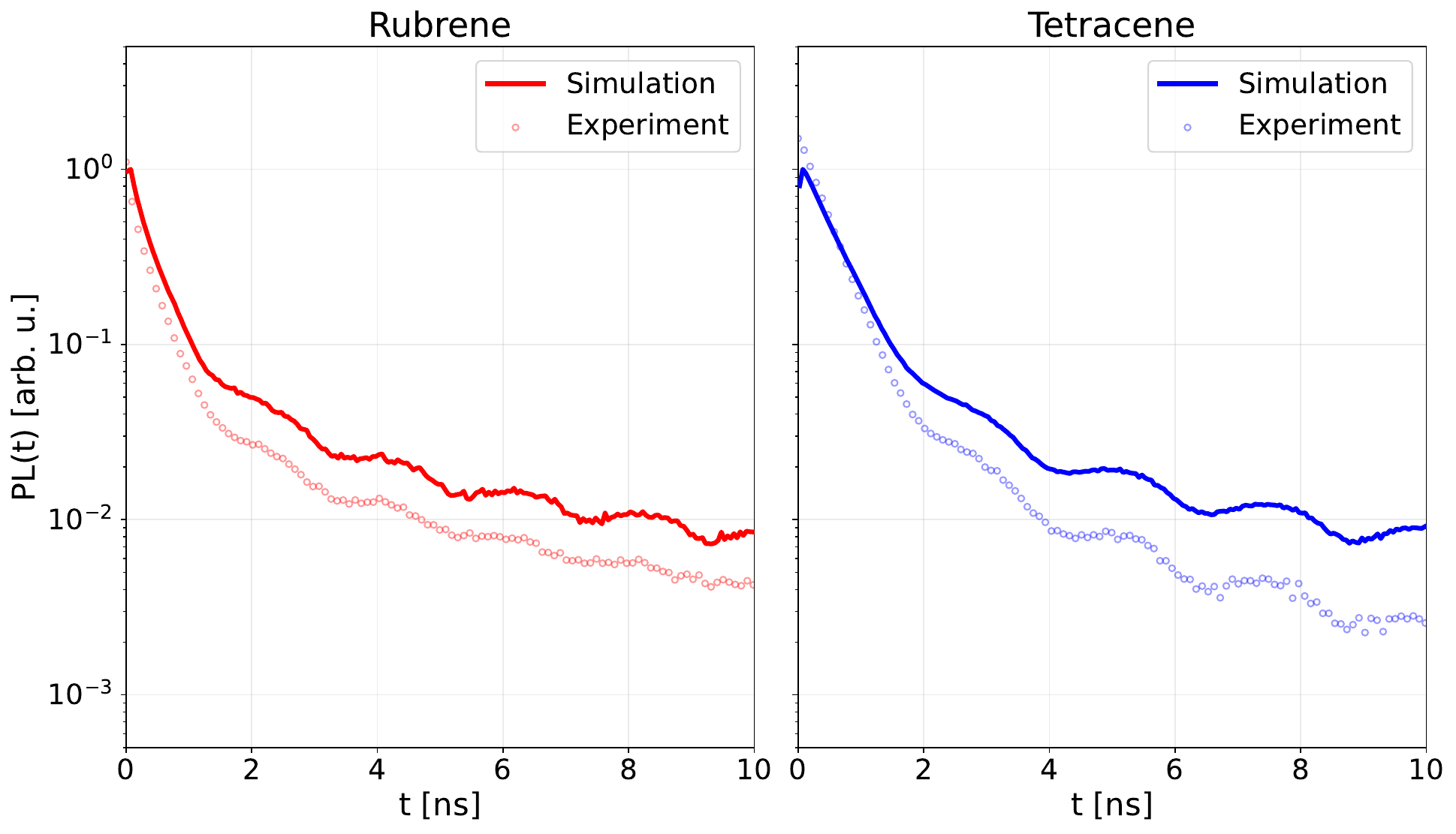}
\caption{ Comparison of experimental and calculated time-resolved photoluminescence for rubrene and tetracene. Open circles are the  experimental data. Thick solid curves are the corresponding theoretical result using the material parameters in Table~\ref{tab:mc_parameters}. Theory and experiment agree for both the overall decay profile and the oscillatory structure of the quantum beats. The  beat frequencies are $f$ = 0.50 GHz for rubrene and $f$ = 0.43 GHz for tetracene, chosen by appropriate orientation of the magnetic field so that the two data sets are similar. In both cases,  $N_{\rm iter}$ = \num{1.0e8}.  The theoretical photoluminescence curves are normalized to unity at their respective maxima. The experimental curves are shifted along the logarithmic vertical axis to visually separate them from each other.}
\label{fig:tdpl_exp_compare} 
\end{figure}

Fig.~\ref{fig:tdpl_exp_compare} also shows a direct representative comparison between experiment and the theoretical prediction obtained using the rubrene and tetracene parameters given in Table~\ref{tab:mc_parameters}. The agreement in both the decay envelope and the oscillatory behavior once again demonstrates that our theoretical model and its Monte Carlo implementation capture the essential physics of the photoluminescence quantum beats in an entangled triplet-pair population. In particular, the relative strength and persistence of the oscillations are well reproduced, indicating that the balance between coherent evolution and stochastic  events is correctly described by the chosen parameter sets.

Taken together, the comparison of the theoretical predictions and the experimental data in tetracene and rubrene confirms that our model provides a unified description of quantum beat amplitudes across different dynamical regimes. The key result is that the observable beat amplitude is determined by two independent factors: the number of stochastic repeated fusion/fission events, and the amount of phase accumulated between those events. Transport controls the former, while the singlet fission time controls the latter. In both tetracene and rubrene, the quantum beat amplitudes are of the order of a few percent at a frequency of $\sim 1$~GHz. But the root cause for this value  is anisotropic transport for rubrene, and a long fission time for tetracene.

We can also note that in general a short singlet fission time (of the order of just a few percent of the expected quantum beat period) is a precondition for  obtaining large-amplitude quantum beats. But in addition to this, for materials to exhibit large quantum beats it is also necessary that the triplet-pair Hamiltonian remains unchanged during triplet exciton diffusion, or that diffusion hopping times are  as short as the fission time.

\section{Conclusion}

We presented a theoretical framework for understanding quantum beat amplitudes in systems undergoing singlet fission and triplet fusion. 
Three key parameters---the second largest two-dimensional hopping time $\tau_2$, the singlet-fission time, and the fusion probability---play distinct and complementary roles in determining the quantum beat amplitude. 

Our analysis showed that the quantum beat amplitudes in rubrene and tetracene are governed by different physical mechanisms. In rubrene, the amplitude is primarily controlled by strongly anisotropic transport, whereas in tetracene it is largely determined by the longer singlet-fission timescale. Using the same theoretical framework, we extracted a hopping time of $200 - 300\space$~ps in rubrene and the singlet-fission time of  $150 - 250\space$~ps in tetracene and found values consistent with those obtained previously through independent experimental methods.  

In general, we showed that  quantum-beat amplitudes are not determined solely by the spin Hamiltonian of the triplet pair. Instead, they emerge from the combined effects of coherent spin evolution, transport anisotropy, encounter statistics, and singlet-fission dynamics. 
Understanding and controlling the interplay between these factors provides a route toward using quantum beat amplitudes as a quantitative probe of microscopic exciton dynamics in molecular materials, and establishes a general framework for investigating coherence in singlet-fission systems.

\begin{acknowledgments}
Research supported by the U.S.~Department of Energy, Office of Basic Energy Sciences, Division of Materials Sciences and Engineering, under Award No.~DE-SC0020981.

\end{acknowledgments}

\bibliography{references}

@article{Curran2025Erratum,
	author = {Curran, Gerald and Rex, Zachary and Xallan Wilson, Casper and Weaver, Luke J. and Biaggio, Ivan},
	doi = {10.1103/6vzz-8zys},
	issue = {18},
	journal = {Phys. Rev. Lett.},
	month = {Oct},
	numpages = {1},
	pages = {189901},
	publisher = {American Physical Society},
	title = {Erratum: Persistence of Spin Coherence in a Crystalline Environment [Phys. Rev. Lett. 133, 056901 (2024)]},
	url = {https://link.aps.org/doi/10.1103/6vzz-8zys},
	volume = {135},
	year = {2025}}

@article{Thompson2015_MagneticFieldSF,
	author = {Thompson, N. J. and Congreve, D. N. and Goldberg, D. and Menon, V. M. and Baldo, M. A.},
	doi = {10.1038/nchem.2327},
	journal = {Nature Chemistry},
	pages = {843--849},
	title = {Magnetic field dependence of singlet fission in tetracene},
	volume = {7},
	year = {2015}}

@article{Wan2015_CooperativeExcitonTransport,
	author = {Wan, Yan and Guo, Zhi and Zhu, Tong and Yan, Suxia and Johnson, Justin and Huang, Libai},
	doi = {10.1038/nchem.2320},
	journal = {Nature Chemistry},
	pages = {785--792},
	title = {Cooperative singlet and triplet exciton transport in tetracene crystals visualized by ultrafast microscopy},
	volume = {7},
	year = {2015}}

@article{Chabr1981_QuantumBeatsTetracene,
	author = {Chabr, M. and Wild, U. P. and F\"unschilling, J. and Zschokke-Gr\"anacher, I.},
	doi = {10.1016/0301-0104(81)80221-2},
	journal = {Chemical Physics},
	number = {3},
	pages = {425--435},
	title = {Quantum beats of prompt fluorescence in tetracene crystals},
	volume = {57},
	year = {1981}}

@article{Wilson2013TetraceneFission,
	author = {Wilson, Mark W. B. and Rao, Akshay and Johnson, Kerr and G\'elinas, Simon and di Pietro, Riccardo and Clark, Jenny and Friend, Richard H.},
	doi = {10.1021/ja4084622},
	journal = {Journal of the American Chemical Society},
	number = {44},
	pages = {16680--16688},
	title = {Temperature-Independent Singlet Exciton Fission in Tetracene},
	volume = {135},
	year = {2013}}

@article{Curran2025TransportDecoherence,
	author = {Curran, Gerald III and Weaver, Luke J. and Rex, Zachary and Biaggio, Ivan},
	doi = {10.1103/PhysRevB.112.214305},
	journal = {Physical Review B},
	pages = {214305},
	title = {Transport-induced decoherence of the entangled triplet exciton pair},
	volume = {112},
	year = {2025}}

@article{Felter2019_SingletFissionCrystals,
	author = {Felter, Kevin M. and Grozema, Ferdinand C.},
	doi = {10.1021/acs.jpclett.9b00754},
	journal = {Journal of Physical Chemistry Letters},
	number = {22},
	pages = {7208--7214},
	title = {Singlet Fission in Crystalline Organic Materials: Recent Insights and Future Directions},
	volume = {10},
	year = {2019}}

@article{Akselrod2014,
	author = {Akselrod, Gleb M. and Deotare, Parag B. and Thompson, Nicholas J. and Lee, Jiye and Tisdale, William A. and Baldo, Marc A. and Menon, Vinod M. and Bulovi{\'c}, Vladimir},
	doi = {10.1038/ncomms4646},
	journal = {Nature Communications},
	number = {3646},
	pages = {1--9},
	title = {Visualization of exciton transport in ordered and disordered molecular solids},
	volume = {5},
	year = {2014}}

@article{Alfano1973,
	author = {Alfano, Robert R. and Shapiro, S. L. and Pope, M.},
	journal = {Optics Communications},
	pages = {388--391},
	title = {Fission Rate of Singlet Excitons in a Tetracene Single Crystal Measured with Picosecond Laser Pulses},
	volume = {9},
	year = {1973}}

@article{Burdett2013,
	author = {Burdett, Jonathan J. and Bardeen, Christopher J.},
	journal = {Accounts of Chemical Research},
	number = {6},
	pages = {1312-1320},
	title = {The Dynamics of Singlet Fission in Crystalline Tetracene and Covalent Analogs},
	volume = {46},
	year = {2013}}

@article{Finton2019,
	author = {Finton, Drew M. and Wolf, Eric A. and Zoutenbier, Vincent S. and Ward, Kebra A. and Biaggio, Ivan},
	journal = {AIP Advances},
	month = {09},
	number = {9},
	pages = {095027},
	title = {{Routes to singlet exciton fission in rubrene crystals and amorphous films}},
	volume = {9},
	year = {2019}}

@article{Wolf2018,
	author = {Wolf, Eric A. and Finton, Drew M. and Zoutenbier, Vincent and Biaggio, Ivan},
	journal = {Applied Physics Letters},
	number = {8},
	pages = {083301},
	title = {{Quantum beats of a multiexciton state in rubrene single crystals}},
	volume = {112},
	year = {2018}}

@article{Burdett2012,
	author = {Burdett, Jonathan J. and Bardeen, Christopher J.},
	journal = {Journal of the American Chemical Society},
	month = {05},
	number = {20},
	pages = {8597--8607},
	title = {Quantum Beats in Crystalline Tetracene Delayed Fluorescence Due to Triplet Pair Coherences Produced by Direct Singlet Fission},
	volume = {134},
	year = {2012}}

@article{Curran2024,
	author = {Curran, Gerald and Rex, Zachary and Xallan Wilson, Casper and Weaver, Luke J. and Biaggio, Ivan},
	issue = {5},
	journal = {Phys. Rev. Lett.},
	numpages = {5},
	pages = {056901},
	title = {Persistence of Spin Coherence in a Crystalline Environment},
	volume = {133},
	year = {2024}}

@article{Wolf2021,
	author = {Wolf, Eric A. and Biaggio, Ivan},
	journal = {Physical Review B},
	month = {05},
	number = {20},
	pages = {L201201--},
	title = {Geminate exciton fusion fluorescence as a probe of triplet exciton transport after singlet fission},
	volume = {103},
	year = {2021}}

@article{Merrifield1968,
	author = {Merrifield, R. E.},
	doi = {10.1063/1.1669777},
	journal = {J. Chem. Phys.},
	pages = {4318--4319},
	title = {Theory of Magnetic Field Effects on the Mutual Annihilation of Triplet Excitons},
	volume = {48},
	year = {1968}}

@article{Irkhin2011,
	author = {Irkhin, Pavel and Biaggio, Ivan},
	journal = {Physical Review Letters},
	month = {07},
	number = {1},
	pages = {017402--},
	title = {Direct Imaging of Anisotropic Exciton Diffusion and Triplet Diffusion Length in Rubrene Single Crystals},
	volume = {107},
	year = {2011}}

@article{Lim2004Exciton,
	author = {Lim, Sang-Hyun and Bjorklund, Thomas G. and Spano, Frank C. and Bardeen, Christopher J.},
	doi = {10.1103/PhysRevLett.92.107402},
	journal = {Physical Review Letters},
	number = {10},
	pages = {107402},
	title = {Exciton Delocalization and Superradiance in Tetracene Thin Films and Nanoaggregates},
	volume = {92},
	year = {2004}}

@article{Suna1970,
	author = {Suna, A.},
	doi = {10.1103/PhysRevB.1.1716},
	journal = {Physical Review B},
	month = {Feb},
	number = {4},
	pages = {1716--1739},
	title = {Kinematics of Exciton-Exciton Annihilation in Molecular Crystals},
	volume = {1},
	year = {1970}}

@article{Funfschilling1985,
	author = {F{\"u}nfschilling, J. and Zschokke-Gr{\"a}nacher, I.},
	journal = {Helvetica Physica Acta},
	pages = {358--365},
	title = {Quantum Beats in the Fluorescence Decay of Tetracene Crystals},
	volume = {58},
	year = {1985}}

@article{Irkhin2012Rubrene,
	author = {Irkhin, Pavel and Ryasnyanskiy, Aleksandr and Koehler, Marlus and Biaggio, Ivan},
	doi = {10.1103/PhysRevB.86.085143},
	journal = {Physical Review B},
	month = {August},
	pages = {085143},
	publisher = {American Physical Society},
	title = {Absorption and photoluminescence spectroscopy of rubrene single crystals},
	volume = {86},
	year = {2012}}

@article{Piland2015Tetracene,
	author = {Piland, Geoffrey B. and Bardeen, Christopher J.},
	doi = {10.1021/acs.jpclett.5b00569},
	journal = {The Journal of Physical Chemistry Letters},
	number = {10},
	pages = {1841--1846},
	publisher = {American Chemical Society},
	title = {How Morphology Affects Singlet Fission in Crystalline Tetracene},
	volume = {6},
	year = {2015}}

@article{Biaggio2026,
	author = {Biaggio, Ivan},
	doi = {10.1039/d6tc02369j},
	earlyaccessdate = {SEP 2026},
	eissn = {2050-7534},
	issn = {2050-7526},
	journal = {Journal of Materials Chemistry C},
	month = {OCT 1},
	number = {37},
	pages = {17106 - 17133},
	title = {What rubrene taught us: from optical absorption and emission to exciton dynamics in a model organic semiconductor},
	volume = {14},
	year = {2026}}

@article{Smith1972Tetracene,
	author = {Smith, A. W. and Weiss, C.},
	doi = {10.1016/0009-2614(72)80250-1},
	journal = {Chemical Physics Letters},
	number = {4},
	pages = {507--511},
	title = {Fluorescence Decay Time Measurements in Tetracene Crystals},
	volume = {14},
	year = {1972}}

@article{Fleming1977Tetracene,
	author = {Fleming, G. R. and Millar, D. P. and Morris, G. C. and Morris, J. M. and Robinson, G. W.},
	journal = {Australian Journal of Chemistry},
	pages = {2353--2359},
	title = {Exciton Fission and Annihilation in Crystalline Tetracene},
	volume = {30},
	year = {1977}}

@article{Klein1978TripletPairs,
	author = {Klein, G.},
	journal = {Chemical Physics Letters},
	pages = {202--206},
	title = {Kinematics of Triplet Pairs in Anthracene and Tetracene Crystals},
	volume = {57},
	year = {1978}}

@article{Robertson1961Tetracene,
	author = {Robertson, J. Monteath and Sinclair, V. C. and Trotter, J.},
	doi = {10.1107/S0365110X61002023},
	journal = {Acta Crystallographica},
	number = {7},
	pages = {697--704},
	title = {The Crystal and Molecular Structure of Tetracene},
	url = {https://doi.org/10.1107/S0365110X61002023},
	volume = {14},
	year = {1961}}
\bibliographystyle{unsrt}

\end{document}